# On-Chip Single-Photon Sifter

Ali W. Elshaari[†*], Iman Esmaeil Zadeh[‡*], Andreas Fognini[‡], Michael E. Reimer[§], Dan Dalacu[∥], Philip J. Poole[∥], Val Zwiller[†], Klaus D. Jöns[†]

[†] Quantum Nano Photonics Group, Department of Applied Physics and Center for Quantum Materials, Royal Institute of Technology (KTH), Stockholm 106 91, Sweden

[‡] Kavli Institute of Nanoscience Delft, Delft University of Technology, Delft 2628 CJ, The Netherlands

[§] Institute for Quantum Computing and Department of Electrical & Computer Engineering, University of Waterloo, Waterloo, ON N2L 3G1, Canada

[∥] National Research Council of Canada, Ottawa, ON K1A 0R6, Canada

* Equal contribution

**Abstract:** Quantum states of light play a pivotal role in modern science[1] and future photonic applications[2]. While impressive progress has been made in their generation and manipulation with high fidelities, the common table-top approach is reaching its limits for practical quantum applications. Since the advent of integrated quantum nanophotonics[3] different material platforms based on III-V nanostructures-, color centers-, and nonlinear waveguides[4-8] as on-chip light sources have been investigated. Each platform has unique advantages and limitations in terms of source properties, optical circuit complexity, and scaling potentials. However, all implementations face major challenges with efficient and tunable filtering of individual quantum states[4], scalable integration and deterministic multiplexing of on-demand selected quantum emitters[9], and on-chip excitation-suppression[10]. Here we overcome all of these challenges with a novel hybrid and scalable nanofabrication approach to generate quantum light on-chip, where selected single III-V quantum emitters are positioned and deterministically integrated in a CMOS compatible circuit[11] with controlled on-chip filtering and excitation-suppression.

Furthermore, we demonstrate novel on-chip quantum wavelength division multiplexing, showing tunable routing of single-photons. Our reconfigurable quantum photonic circuits with a foot print one million times smaller than similar table-top approaches, offering outstanding excitation suppression of more than 95 dB and efficient routing of single photons over a bandwidth of 40 nm, are essential to unleash integrated quantum optical technologies' full potential.

The implementation of the linear optics quantum computation scheme[12] requires large resources and remarkable photon detection and generation efficiencies[13]. Thus, the progress of quantum information processing[14] and sensing implementations[15] using quantum states of light strongly depends on miniaturization and simultaneous integration of the main elements of a quantum circuit, i.e., sources, optical gates, and detectors[16]. Photonic integration provides a means of miniaturization to efficiently integrate elements of the quantum circuit on-chip with low insertion losses, high efficiencies[11, 17], and high density. Generally, the type of application determines which photonic platform is used, but there are some universal requirements in a quantum photonic circuit regardless. These requirements are: deterministic integration of multiple on-demand selected single-photon sources[11, 18], demonstration of complex photonic circuits for qubit manipulation[6, 19, 20], and on-chip detection[17]. The key ingredient to realize such a complete system is on-chip spectral filtering and multiplexing of multiple quantum emitters. Several reports addressed single photon filtering and on-chip excitation suppression [6, 19, 21], but were limited to quantum emitters of probabilistic nature in nonlinear optical waveguides. In contrast, semiconductor quantum dot (QD) single-photon sources[22] are suitable for on-demand photon generation[23], but there are no reports of on-chip filtering and multiplexing of QD single-photon sources. QD-based single-photon sources have an additional advantage that they also offer the possibility for on-chip electrical excitation[24] and wavelength tunable entangled photon emission[25, 26]. In conjunction with two-photon interference visibilities approaching unity[27] and high photon flux rates[28, 29], quantum dots are therefore attractive for on-chip quantum optical applications. However, there are challenges to realize III-V quantum photonic circuits [30-32]. These challenges include deterministic integration of selected QDs into optical waveguides/cavities, efficient filtering of specific quantum transitions within the emission spectrum, on-chip pump-suppression, and multiplexing of multiple QDs. In this letter, we overcome all of these challenges by realizing novel integrated quantum photonic circuits. Our circuits generate quantum light on-chip relying on a fully-integrated and tunable spectral filtering to isolate optically-active QD transitions. Moreover, to emphasize the controlled nature of the process and the vast range of new applications it enables, we demonstrate a novel on-chip Quantum Wavelength Division Multiplexing (QWDM) channel with tunable routing.

**Figure.1a** shows an artistic view of the hybrid quantum photonic circuit used in our work comprising of a nanowire-based quantum light source embedded in a photonic waveguide with tunable ring resonator filter. The quantum emitter consists of an InAsP QD embedded in a pure wurtzite InP nanowire. Details on the nanowire growth can be found in Ref. [33]. After pre-characterizing the nanowire QDs on the growth chip, we select emitters based on their emission wavelength, brightness, and line width,  and then transfer them to the desired location using a custom-built nanomanipulation tool[11]. Next, the photonic channel is carefully designed to encapsulate the transferred nanowire-based quantum emitter. The photonic channel consists of a 200 nm thick Silicon Nitride (SiN) layer processed into a 800 nm wide waveguide and cladded with a PMMA layer. **Figures.1b-d** summarize the quantum photonic circuit fabrication process where we first deterministically position the pre-selected nanowire-based quantum emitter and then construct the photonic waveguide around it. A detailed description of the fabrication steps and nanowire nanomanipulation can be found in the **Methods** section. **Figure.1e** shows a microscope image of a nanowire-based quantum emitter integrated in a photonic waveguide. The nanowire is encapsulated in the SiN waveguide, allowing for ease of excitation and collection in the forward and backward directions with a coupling efficiency of approximately 24% in the forward and backward direction [11]. **Figure.1f** and **Figure.1g** show the collected photoluminescence (PL) in the forward and backward directions of the photonic channel, respectively. If desired this configuration can directly act as an on-chip non-polarizing beamsplitter and the observed ¾ ratio between the forward and backward directions can be engineered at will. The ratio mainly depends on the nanowire shape [11] and differences in the out-of-chip coupling and propagation losses of the two waveguides (see **Methods**).

As shown in **Figure.2a,** the emitted photons from the nanowire QD in the forward direction are routed to an electrically tunable 70 µm-radius ring resonator. The SiN waveguide width, height, and 180 nm separation between the ring resonator and the bus waveguides are designed to achieve critical coupling for TE modes (See **Supplementary Note S.1** for TM transmission). **Figure.2b** and **Figure.2c** present the measured through-port and drop-port transmission of the ring resonator TE mode (electric field parallel to the substrate), respectively. The free spectral range of the resonator is approximately 0.96 nm at the quantum dot emission wavelength with a resonance FWHM of 0.13 nm. To efficiently tune the ring resonator to the QD emission, the top cladding was carefully chosen to have a negative thermo-optic coefficient, approximately 10 times larger than the positive thermo-optic coefficient of the SiN waveguide core [34, 35]. **Figure.2d** shows the ring resonances shift to shorter wavelength (higher energies) with increasing voltage. In contrast, as shown in **Figure.2e,** the QD emission shifts in the opposite direction to longer wavelengths (lower energies) for increasing voltage. This counteracting tuning mechanism of the resonator and the QD enhances the tuning rate. To minimize the thermal coupling between the nanowire and the tunable filter, the PMMA cladding was removed from most of the substrate except for small areas surrounding the photonic channels to preserve optical mode confinement. The voltage range between 0 and 15 V covers already 120% of the total free spectral range of the resonator, we therefore can efficiently select any spectral line in a range of 40 nm without degradation in the performance of the ring resonator. We independently studied the QD emission as a function of chip temperature to estimate the actual temperature of the QD when a certain voltage is applied to the ring resonator filter, see **Supplementary Note S.2** for measurement details.

As shown in **Figure.2f** and **Figure.2g**, specific resonant modes of the resonator can be tuned to either the trion (T) or exciton (X) transition of the QD. Thus, single-photons are routed from the through-port to the drop-port with more than 15 dB selectivity measured at the through-port. Careful engineering of the resonator allows for selective filtering based on wavelength and/or polarization. This capability becomes important in entanglement type experiments with QDs utilizing the biexciton-exciton cascade. Finally, the versatile nature of SiN photonics allows for a vast range of more complex multi-staged circuitry[36], with third order nonlinearity that can be employed to perform on-chip inter- and intra-band single-photon wavelength conversion [37].

Next, we generate quantum light on-chip without the use of any bulky or lossy external wavelength filters. The experimental setup is shown in **Figure.3a**. The QD is excited from the top using a CW 532 nm laser (see **Methods** for more information about QD excitation and photon collection). The collected emission from the waveguide facet is either filtered through a monochromator, then detected by an Avalanche Photodiode (APD) (Case 1), or it can be directly detected by the APDs (Case 2). **Figure.3b** shows the unfiltered emission spectra of the encapsulated nanowire QD. Both the exciton and trion transition are visible. **Figure.3c** presents a high resolution measurement performed on the drop-port emission of the ring resonator over a wide wavelength range (from 500 nm to 950 nm) after the filter voltage is tuned to align the ring resonator resonance to the QD trion transition. The inset focuses on the filtered trion transition, showing the full suppression of the exciton line. Despite the presence of an intense laser excitation of power density ~0.5µW/µm$^2$ from the top, in addition to the exciton and bulk InP nanowire emission, only the selected trion emission line is collected. With our filtering, we achieve an on-chip excitation suppression of more than 95 dB. (see **Supplementary Note S.3** for detailed information about excellent on-chip excitation suppression).

In order to characterize the on-chip ring resonator filter we bench-mark its performance against a commercial table-top monochromator, an Acton 750 mm with 1800 grooves per millimeter grating. The bandwidth of the monochromator is approximately 0.02 nm with an overall transmission of ~25% and peak attenuation in the rejection band of > 80 dB. We perform second-order correlation measurements on the trion line in both cases: case 1 – using the monochromator as a filter and case 2 – bypassing the monochromator and using the on-chip ring resonator filter. Case 1 is shown in **Figure.3e** where the multi-photon probability at zero-time delay is g$^2$(0) = 0.13 ± 0.04 (see **Supplementary Note S.4** for measurement details and cross-correlation measurements between the trion and the exciton lines). The result for case 2 with only on-chip filtering is shown in **Figure.3d**. In the latter case, the multi-photon probability at zero-time delay is g$^2$(0) = 0.40 ± 0.05, demonstrating the emission from a single quantum emitter on chip. As a final check, the measurement is repeated without the use of any filtering, shown in **Figure.3f**. As expected, without the monochromator or the ring resonator filtering we observe no signature of non-classical light emission. Comparing the previous measurements clearly shows the functionality of our novel on-chip single-photon filtering using a single-stage ring resonator. The unwanted coupled light to the drop port can be further reduced by coupled resonators with larger free spectral range or narrower bandwidth [36].

Having demonstrated the main building blocks of our hybrid quantum photonic circuit, namely quantum emitter integration and filtering of individual exciton transitions, we move to more complex architectures. To increase the on-chip qubit data rate, advanced multiplexing techniques can be adopted from classical fiber-optic communication systems like wavelength-division-multiplexing. However, up to now the missing on-chip filtering hindered the realization of QWDM transmission. Using our elegant tunable ring resonator filter we are able to perform this task. **Figure.4** shows a schematic view of a fabricated 2-qubit on-chip QWDM channel. For this device the nanowires were butt-coupled to the photonic waveguide (see **Supplementary Note S.5** for simulation of nanowire-waveguide coupling). Details of the device fabrication and nanomanipulation of the individual nanowires are included in the **Methods** section. The two QDs can be excited independently or simultaneously to create a quantum multi-wavelength integrated channel. Afterwards the photons will be selectively decoupled from the main transmission waveguide based on their wavelength. **Figure.4a** shows the collected emission spectrum from the through-port waveguide. The spectrum consists of a wavelength-multiplexed packet carrying the emission of two QDs, labeled QD1 and QD2, separated by ~10 nm. Demultiplexing the emission is performed using the on-chip tunable ring resonator filter, which only couples the resonant optical modes to the drop-port of the photonic circuit. As shown in **Figure.4b**, specific transitions of QD1 and QD2 are routed deterministically to the drop-port waveguide as a function of the on-chip filter tuning voltage. In **Figure.4c** we present the integrated intensity of QD1 and QD2 in the drop-port as a function of voltage. As we tune the ring resonator, the intensity of QD1 and QD2 follow a Lorentzian shape given by the transmission function of the ring resonator filter. As we increase the voltage further, the QD1 peak photoluminescence is decoupled from the drop-port and we couple the emission of the QD2 peak. See **Supplementary Note S.6** for more QWDM measurements. Our device can be extended to incorporate more emitters and pulsed excitation to realize temporal and WDM quantum link on-chip[38]. With the recent advances of generating entangled photon pairs from similar nanowire QDs [39, 40], it is possible to multiplex/demultiplex entangled photon pairs in complex network architectures.

In summary, we have deterministically integrated quantum emitters on a CMOS compatible platform with tunable ring resonator filters that show ultra-efficient pump rejection. This allowed us to demonstrate a novel reconfigurable 2-qubit QWDM channel which can be extended to include more quantum emitters and incorporate additional coding schemes.  By eliminating the need for off-chip components, our results open up new possibilities for large-scale quantum photonic systems with on-chip single- and entangled- photon sources. Our approach, resulting in efficient tunable filtering and routing of single-photons, allows for the integration of complex architectures and will enable experiments far beyond two particles. It brings us an important step closer to realize the ambitious schemes of (linear) optical quantum computing that have been put forward over the past decades.

# Methods:

**Nanomanipulator:** The tool is capable of positioning and aligning nano-sized objects with less than 250 nm position resolution and < 1 degrees' rotation resolution. It consists of a Tungsten tip mounted on a xyz high precision differential-stage, all integrated in a high resolution imaging system. Relying on van der Waals forces between the nanowires and the Tungsten tip, nanowires can be selectively transferred from the growth chip to another substrate for further processing. More details about the nanowire transfer process and the nanomanipulator can be found in Ref. [11].

**Photonic circuit fabrication, encapsulation device:** Starting with a bare silicon wafer, 2.4 µm of thermal oxide is formed to serve as the bottom cladding of the waveguide. Using e-beam lithography, metal evaporation and lift-off process, metallic structures including marker fields and heaters (resistance 2.8 k-Ohms) were created on the oxide layer. Next, nanowire QDs were transferred with the nanomanipulation tool to the substrate, followed by a deposition of 200 nm of SiN using PECVD process at 300 °C [34]. Waveguides and ring resonators were patterned using 100 keV e-beam lithography on a 950 K PMMA resist. After developing the resist, features were transferred to the SiN by complete etching of the SiN layer using $CHF_3$/Ar based reactive ion etching. This was followed by a short $O_2$ plasma cleaning step. Next, the devices were covered with ~1 µm thick PMMA to provide symmetric mode confinement and negative thermo-optic tuning of the rings. Finally, to reduce the thermal coupling between the heaters and the nanowires, and to provide easy access to the electrical contacts for bonding, the PMMA layer was removed from the majority of the substrate, except in regions surrounding the photonic circuit.

**Photonic circuits fabrication, butt coupling devices:** The process is similar to the encapsulation devices for heaters (resistance 0.9 k-Ohms) and photonic elements fabrication, except that nanowires are transferred after the circuit is fabricated. After cladding the whole chip with PMMA, an additional e-beam lithography step was performed to form openings in the cladding at the nanowire-waveguide coupler region. Finally, the nanowires are transferred and aligned using the nanomanipulation tool. The approach provides more flexibility in which the quantum sources are transferred after the photonic circuits are fabricated and tested, this enables more control and selectivity in matching the photonic circuits with the quantum emitters.

**QD excitation and photon collection:** The chip design employed a U-shape structure [34] with input and output waveguide separation of 40 μm. This simplifies coupling to/from the side facet of the chip and separates input and output beams spatially with a two foci setup. The chip was cleaved along a crystallographic direction to achieve flat/smooth facet for coupling. Photons propagating in the SiN waveguides were collected using a 50X objective with an NA of 0.75. The QDs can be either excited in-plane through the waveguide using a HeNe laser (632 nm), or out-of-plane using a solid-state green laser (532 nm).

## Acknowledgement:

I. EZ. And K.D.J acknowledge Andrea Pescalini for initial scientific discussions. A.W.E and V.Z acknowledge the support of ERC consolidator grant (ERC-2012-StG) and VR grant for international recruitment of leading researchers (Ref: 2013-7152). I.EZ acknowledges the support of Dutch Foundation for Fundamental Research on Matter (FOM projectruimte 10NQO02) and support from Single Quantum B.V. (SQ). M.E.R. acknowledges support from Industry Canada. K.D.J. acknowledges funding from the MARIE SKŁODOWSKA-CURIE Individual Fellowship under REA grant agreement No. 661416 (SiPhoN).

## Author contributions:

A.W.E, I. EZ, and K.D.J conceived and designed the experiment. A.W.E and I.EZ designed and fabricated samples and carried out the experiments with the help of K.D.J and A.F.. The data was analyzed by A.W.E, I. EZ, and K.D.J.. D.D and P.J.P. fabricated the nanowire quantum dot. M.E.R contributed to the nanowire-based single-photon sources project. K.D.J. and V.Z led and supervised the project. A.W.E wrote the manuscript with inputs from I.EZ, M.E.R, V.Z., and K.D.J..

## Competing financial interests:

The authors declare no competing financial interests.

# References :

[1] I.A. Walmsley, Science, 348 (2015) 525-530.

[2] B. Korzh, C.C.W. Lim, R. Houlmann, N. Gisin, M.J. Li, D. Nolan, B. Sanguinetti, R. Thew, H. Zbinden, Nat Photon, 9 (2015) 163-168.

[3] J.L. O'Brien, A. Furusawa, J. Vuckovic, Nat Photon, 3 (2009) 687-695.

[4] G. Reithmaier, M. Kaniber, F. Flassig, S. Lichtmannecker, K. Müller, A. Andrejew, J. Vučković, R. Gross, J.J. Finley, Nano Letters, 15 (2015) 5208-5213.

[5] P. Lodahl, S. Mahmoodian, S. Stobbe, Reviews of Modern Physics, 87 (2015) 347-400.

[6] D. Grassani, S. Azzini, M. Liscidini, M. Galli, M.J. Strain, M. Sorel, J.E. Sipe, D. Bajoni, Optica, 2 (2015) 88-94.

[7] M.D. Eisaman, J. Fan, A. Migdall, S.V. Polyakov, Review of Scientific Instruments, 82 (2011) 071101.

[8] I. Aharonovich, A.D. Greentree, S. Prawer, Nat Photon, 5 (2011) 397-405.

[9] A. Laucht, S. Pütz, T. Günthner, N. Hauke, R. Saive, S. Frédérick, M. Bichler, M.C. Amann, A.W. Holleitner, M. Kaniber, J.J. Finley, Physical Review X, 2 (2012) 011014.

[10] A. Faraon, I. Fushman, D. Englund, N. Stoltz, P. Petroff, J. Vuckovic, Nat Phys, 4 (2008) 859-863.

[11] I.E. Zadeh, A.W. Elshaari, K.D. Jöns, A. Fognini, D. Dalacu, P.J. Poole, M.E. Reimer, V. Zwiller, Nano Letters, 16 (2016) 2289-2294.

[12] E. Knill, R. Laflamme, G.J. Milburn, Nature, 409 (2001) 46-52.

[13] P. Kok, W.J. Munro, K. Nemoto, T.C. Ralph, J.P. Dowling, G.J. Milburn, Reviews of Modern Physics, 79 (2007) 135-174.

[14] C.H. Bennett, D.P. DiVincenzo, Nature, 404 (2000) 247-255.

[15] H. Clevenson, M.E. Trusheim, C. Teale, T. Schroder, D. Braje, D. Englund, Nat Phys, 11 (2015) 393-397.

[16] S. Khasminskaya, F. Pyatkov, K. Słowik, S. Ferrari, O. Kahl, V. Kovalyuk, P. Rath, A. Vetter, F. Hennrich, M.M. Kappes, Gol'tsmanG, A. Korneev, C. Rockstuhl, R. Krupke, W.H.P. Pernice, Nat Photon, advance online publication (2016).

[17] W.H.P. Pernice, C. Schuck, O. Minaeva, M. Li, G.N. Goltsman, A.V. Sergienko, H.X. Tang, Nat Commun, 3 (2012) 1325.

[18] Z. Guo-Wei, S. Xiang-Jun, N. Hai-Qiao, Y. Ying, X. Jian-Xing, W. Si-Hang, M. Ben, Z. Li-Chun, N. Zhi-Chuan, Nanotechnology, 26 (2015) 385706.

[19] N.C. Harris, D. Grassani, A. Simbula, M. Pant, M. Galli, T. Baehr-Jones, M. Hochberg, D. Englund, D. Bajoni, C. Galland, Physical Review X, 4 (2014) 041047.

[20] D. Englund, A. Faraon, I. Fushman, N. Stoltz, P. Petroff, J. Vuckovic, Nature, 450 (2007) 857-861.

[21] M.J. Collins, C. Xiong, I.H. Rey, T.D. Vo, J. He, S. Shahnia, C. Reardon, T.F. Krauss, M.J. Steel, A.S. Clark, B.J. Eggleton, Nat Commun, 4 (2013).

[22] P. Michler, A. Kiraz, C. Becher, W.V. Schoenfeld, P.M. Petroff, L. Zhang, E. Hu, A. Imamoglu, Science, 290 (2000) 2282-2285.

[23] M. Muller, S. Bounouar, K.D. Jöns, M. Glässl, P. Michler, Nat Photon, 8 (2014) 224-228.

[24] Z. Yuan, B.E. Kardynal, R.M. Stevenson, A.J. Shields, C.J. Lobo, K. Cooper, N.S. Beattie, D.A. Ritchie, M. Pepper, Science, 295 (2002) 102-105.

[25] Y. Chen, J. Zhang, M. Zopf, K. Jung, Y. Zhang, R. Keil, F. Ding, O.G. Schmidt, Nature Communications, 7 (2016) 10387.

[26] R. Trotta, J. Martin-Sanchez, J.S. Wildmann, G. Piredda, M. Reindl, C. Schimpf, E. Zallo, S. Stroj, J. Edlinger, A. Rastelli, Nat Commun, 7 (2016).

[27] N. Somaschi, V. Giesz, L. De Santis, J.C. Loredo, M.P. Almeida, G. Hornecker, S.L. Portalupi, T. Grange, C. Antón, J. Demory, C. Gómez, I. Sagnes, N.D. Lanzillotti-Kimura, A. Lemaítre, A. Auffeves, A.G. White, L. Lanco, P. Senellart, Nat Photon, 10 (2016) 340-345.

[28] M.E. Reimer, G. Bulgarini, N. Akopian, M. Hocevar, M.B. Bavinck, M.A. Verheijen, E.P.A.M. Bakkers, L.P. Kouwenhoven, V. Zwiller, Nat Commun, 3 (2012) 737.

[29] O. Gazzano, S. Michaelis de Vasconcellos, C. Arnold, A. Nowak, E. Galopin, I. Sagnes, L. Lanco, A. Lemaître, P. Senellart, Nat Commun, 4 (2013) 1425.

[30] N. Prtljaga, R.J. Coles, apos, J. Hara, B. Royall, E. Clarke, A.M. Fox, M.S. Skolnick, Applied Physics Letters, 104 (2014) 231107.

[31] G. Reithmaier, J. Senf, S. Lichtmannecker, T. Reichert, F. Flassig, A. Voss, R. Gross, J.J. Finley, Journal of Applied Physics, 113 (2013) 143507.

[32] K.D. Jöns, U. Rengstl, M. Oster, F. Hargart, M. Heldmaier, S. Bounouar, S.M. Ulrich, M. Jetter, P. Michler, Journal of Physics D: Applied Physics, 48 (2015) 085101.

[33] D. Dalacu, K. Mnaymneh, J. Lapointe, X. Wu, P.J. Poole, G. Bulgarini, V. Zwiller, M.E. Reimer, Nano Letters, 12 (2012) 5919-5923.

[34] A.W. Elshaari, I.E. Zadeh, K.D. Jöns, V. Zwiller, IEEE Photonics Journal, 8 (2016) 1-9.

[35] E. Suhir, Y.C. Lee, C.P. Wong, Micro-and Opto-Electronic Materials and Structures: Physics, Mechanics, Design, Reliability, Packaging: Volume I Materials Physics-Materials Mechanics. Volume II Physical Design-Reliability and Packaging, Springer Science & Business Media2007.

[36] J. Wang, Z. Yao, T. Lei, A.W. Poon, Scientific Reports, 4 (2014) 7528.

[37] Q. Li, M. Davanço, K. Srinivasan, Nat Photon, 10 (2016) 406-414.

[38] C. Xiong, X. Zhang, Z. Liu, M.J. Collins, A. Mahendra, L.G. Helt, M.J. Steel, D.Y. Choi, C.J. Chae, P.H.W. Leong, B.J. Eggleton, Nat Commun, 7 (2016).

[39] M.A.M. Versteegh, M.E. Reimer, K.D. Jöns, D. Dalacu, P.J. Poole, A. Gulinatti, A. Giudice, V. Zwiller, Nat Commun, 5 (2014).

[40] T. Huber, A. Predojević, M. Khoshnegar, D. Dalacu, P.J. Poole, H. Majedi, G. Weihs, Nano Letters, 14 (2014) 7107-7114.

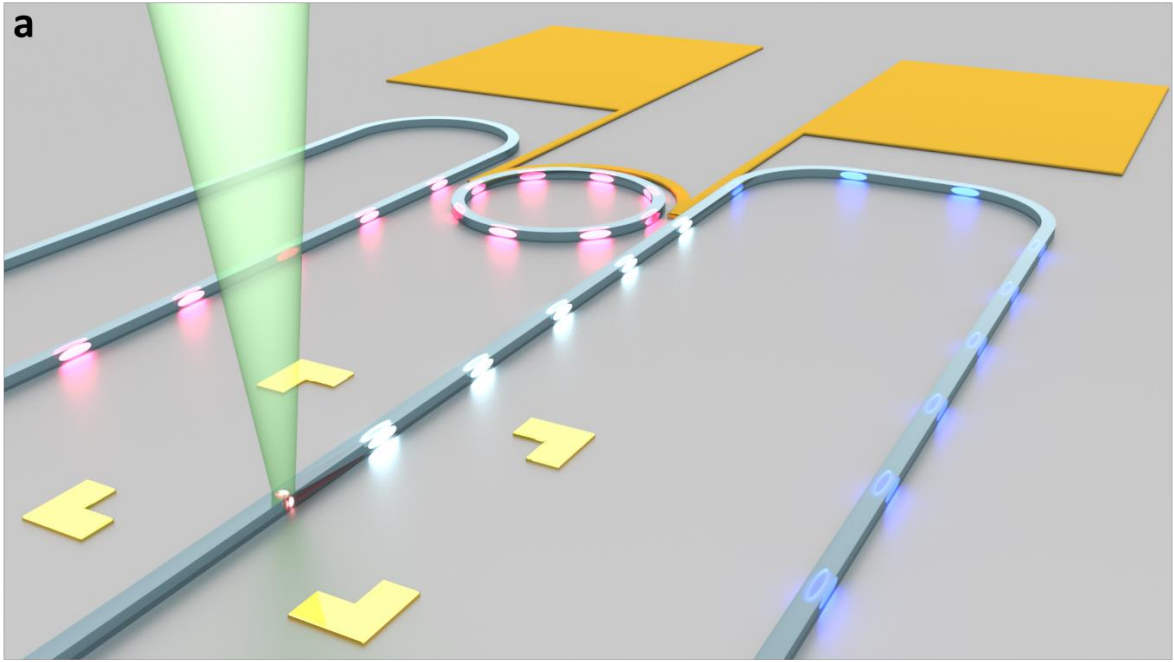

**a**

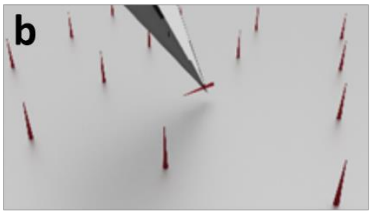
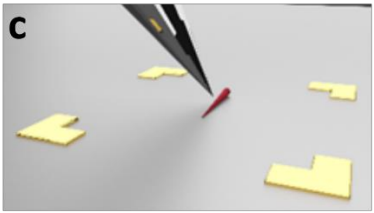
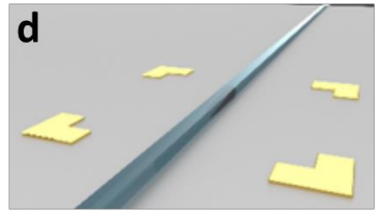

**b** **c** **d**

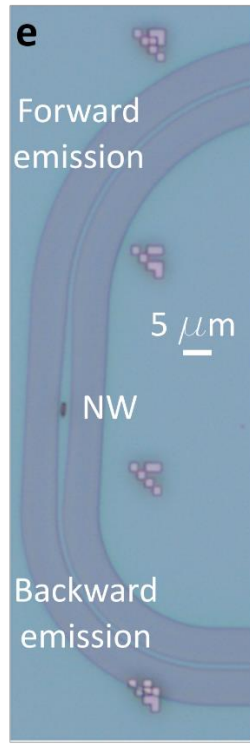

**e**

Forward emission

5 μm

NW

Backward emission

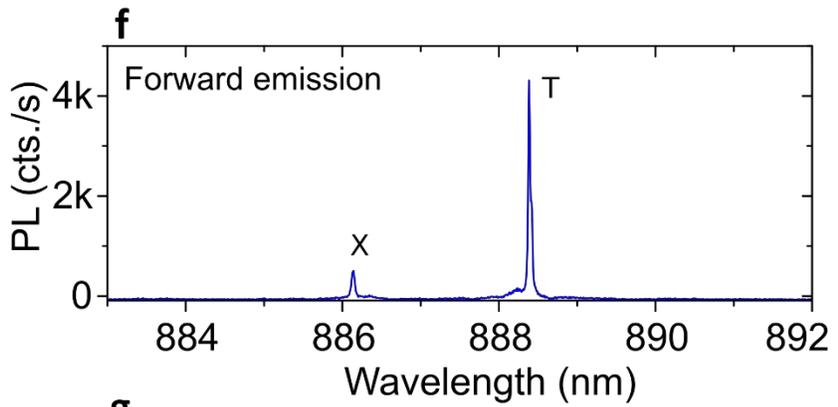

**f**

Forward emission

X

T

PL (cts./s)

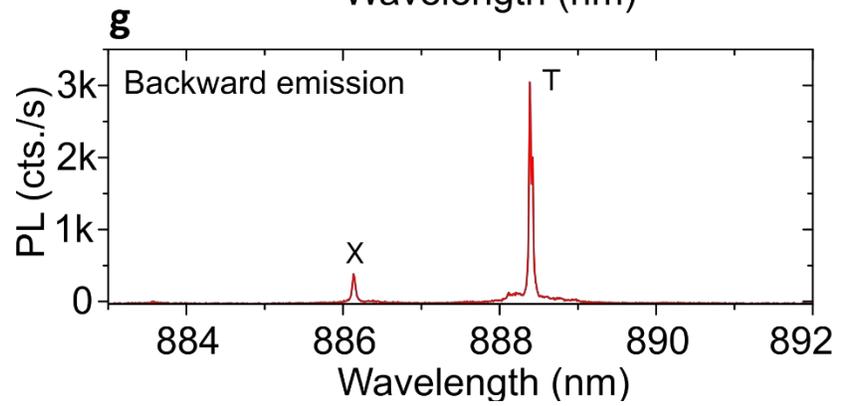

**g**

Backward emission

X

T

PL (cts./s)

**Figure.1** (a) Schematic view of the fabricated hybrid quantum photonic circuit, consisting of an InAsP QD in an InP nanowire (ruby colored) that is integrated with a SiN waveguide (blue) and on-chip tunable ring resonator filter. The ring resonator filter is tuned by applying voltage to the Gold contacts (orange). The out of plane laser (green) excites the QD which emits single photons (ellipsoids) into the waveguide. (b-d) depict the CMOS compatible process-flow for integrating the nanowire–based quantum light sources within the hybrid quantum photonic circuit. Using a custom-built tungsten nanomanipulation tool, the nanowires are transferred from the growth chip to the selected photonic circuit substrate. (e) A microscope image of an integrated single-photon source with a SiN photonic waveguide. Emitted photons are coupled to the SiN photonic channel with the possibility to collect both forward and backward photons independently. The photonic circuits are fabricated with respect to the transferred quantum emitters as described in the **Methods** section. (f) and (g) show the collected forward and backward emission from the nanowire QD, respectively. T and X represent the trion and exciton emission lines, respectively.

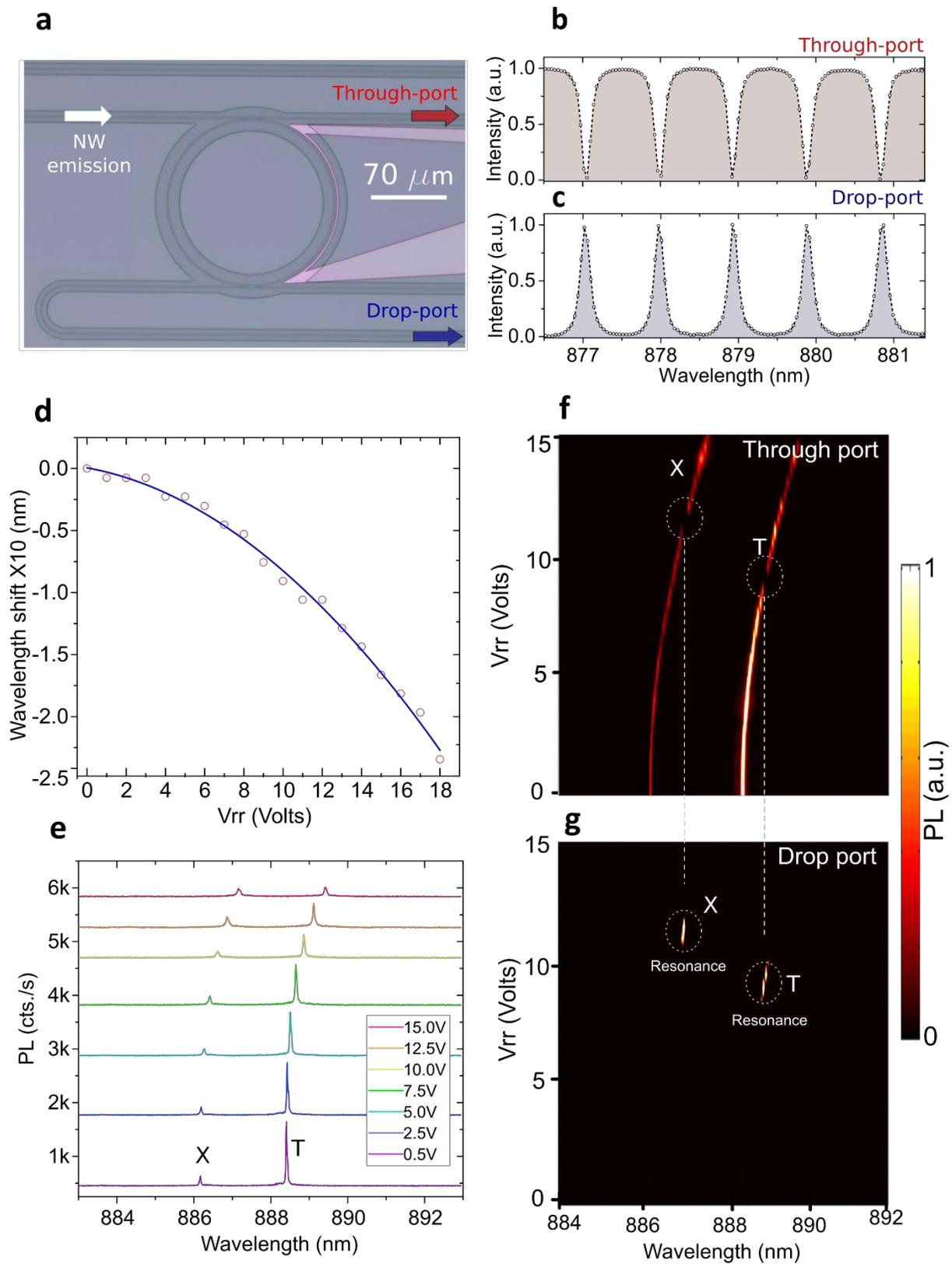

**Figure.2** (a) A microscope image of the ring resonator filter. The filter transmission is controlled with a Titanium resistor. Details of the fabrication process are provided in the **Methods** section. Graphs (b) and (c) show the through-port and drop-port transmission of the ring resonator, respectively. (d) Tuning of the ring resonator filter as a function of the filter voltage (Vrr). The red circles show the measured shifts while the blue line is a fit. The resonances are blue shifted by design, which is achieved by means of the large negative thermo-optic coefficient of the PMMA top cladding. (e) Tuning of the QD transition as a function of the ring resonator voltage. In contrast to the observed shift for the ring resonator, the QD emission red shifts for increasing voltage as expected due to the increase in temperature. Graphs (f) and (g) show the results of selectively routing a unique transition of the QD between the drop-port and through-port of the ring resonator. As we tune the filter resonance using the integrated heater, a single emission line can be tuned in and out of resonance, thus routing single-photons between the through-port and drop-port.

**a**

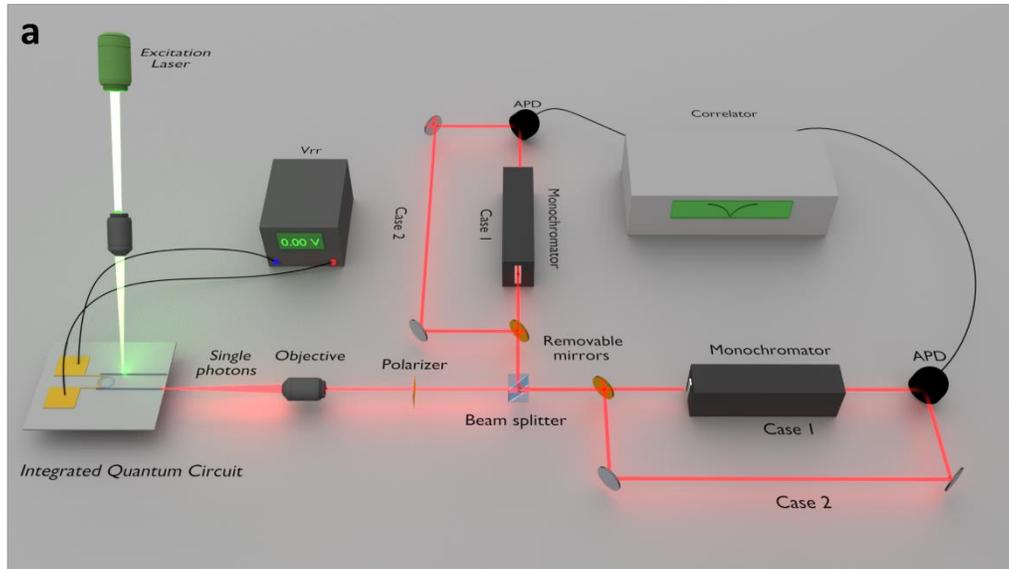

**b**

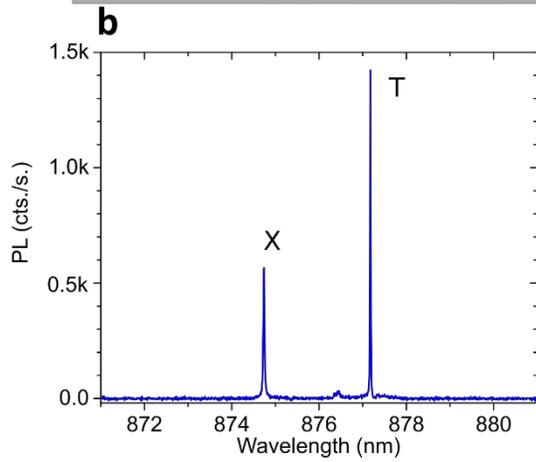

**c**

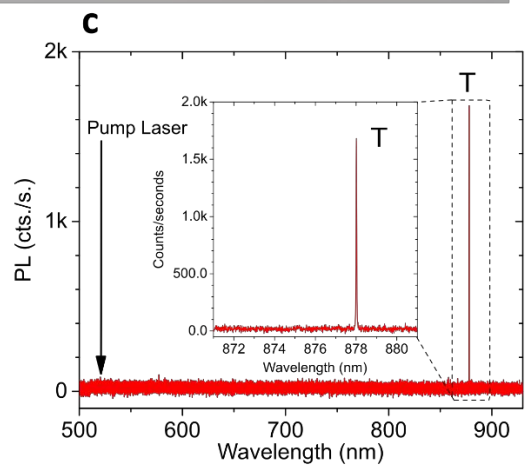

**d**

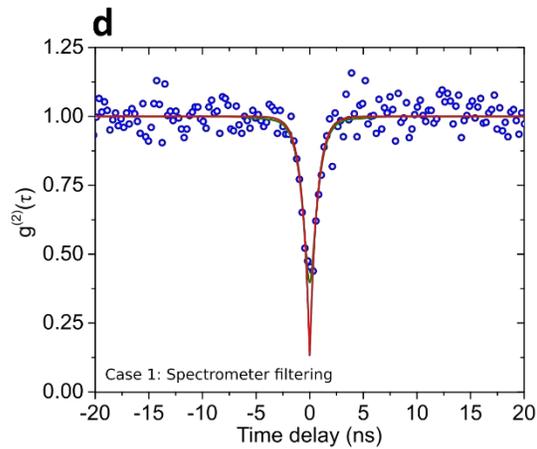

**e**

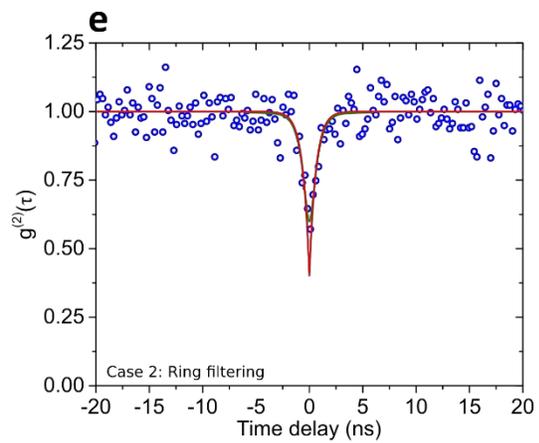

**f**

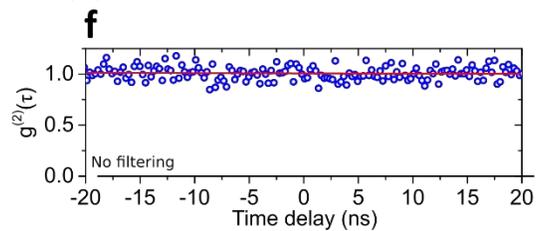

**Figure.3** (a) Experimental setup showing the hybrid integrated quantum circuit with electrical access to control the integrated filters. The setup allows for both in-plane (via the waveguide) and out-of-plane laser excitation. Details of out-of-chip coupling are included in the **Methods** section. The collected emission from the waveguide is either coupled to the APDs after filtering with a monochromator (Case 1), or it can be directly coupled to the APDs with no external wavelength filtering (Case 2). (b) Collected QD emission from the facet of the SiN waveguide in the forward direction. (c) By tuning the on-chip filter, a single QD transition is routed to the drop-port. The inset shows a close-up of the filtered trion (T) emission line. The QD emission wavelength is slightly different in (b) and (c) due to different biases applied to the ring resonator filter. Despite the presence of an intense laser for excitation and InP nanowire emission, the filtered spectrum shows only a single QD transition over a broad wavelength range (500nm to 950nm). (d) Second-order correlation measurement of the QD trion line using an off-chip commercial monochromator for filtering, resulting in a multi-photon probability of $g^2(0) = 0.13 \pm 0.04$ when taking into account the finite temporal resolution of the APDs. (e) Second- order correlation measurement of the QD trion line at the drop-port of the ring resonator after directly coupling it to the APDs. A single stage ring filter is capable of delivering single photons on-chip with multi-photon probability $g^2(0) = 0.41 \pm 0.05$. The results show the excellent performance of the integrated ring resonator filter as compared to the bulky off-chip monochromator. (f) Second-order correlation measurement without any on-chip and off-chip filtering. The results show the expected Poissonian statistics of coherent (uncorrelated) emission. In (d) and (e), the blue circles show the raw data, the green line represents a fit, and the red line represents the fit considering the finite detector response (see **Supplementary Note S.4** for more details).

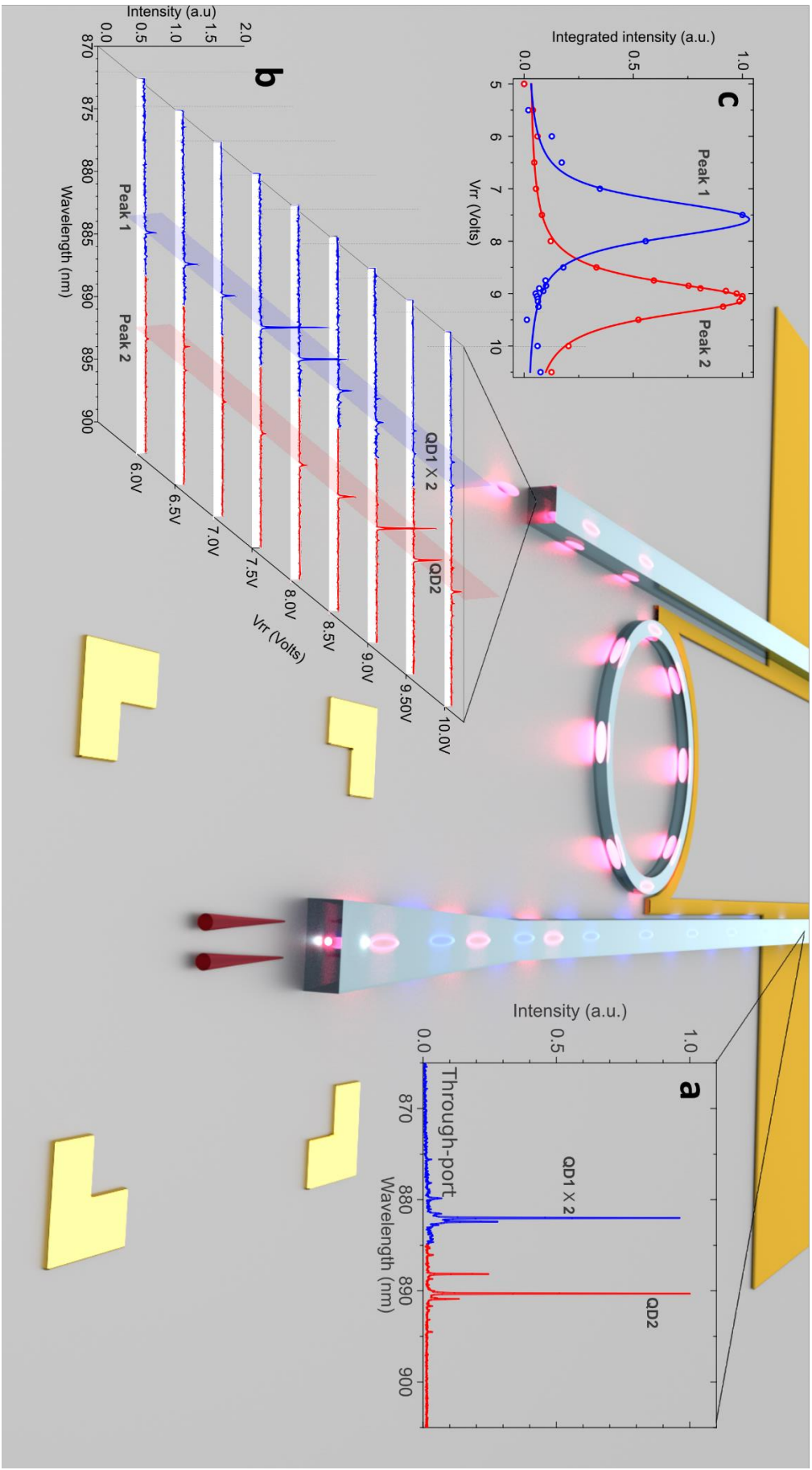

**a**

Through-port

QD1 x 2

QD2

Intensity (a.u.)

Wavelength (nm)

**b**

Peak 1

Peak 2

QD1 x 2

QD2

Intensity (a.u)

Wavelength (nm)

$V_{rr}$ (Volts)

6.0V
6.5V
7.0V
7.5V
8.0V
8.5V
9.0V
9.50V
10.0V

**c**

Peak 1

Peak 2

Integrated intensity (a.u.)

$V_{rr}$ (Volts)

**Figure.4** Artistic image of multiplexing/demultiplexing of two quantum emitters coupled to a photonic circuit with integrated tunable filters. The two nanowires are butt-coupled to a SiN waveguide, each of them emitting photons independently with different colors (depicted as red and blue in this case). The flexibility of the process allows for the possibility of wavelength and modal multiplexing of selected single-photon sources to an already fabricated and characterized photonic circuit, thus making the process highly deterministic. Details of the fabrication and nanowire transfer process are included in the **Methods** section. (a) Collected emission from the through-port waveguide, consisting of a wavelength-multiplexed signal from QD1 and QD2. The spectra are highlighted in red and blue to indicate the individual emission from QD1 and QD2. (b) Selected excitonic transitions of QD1 and QD2 are filtered deterministically to the drop-port waveguide as a function of the on-chip filter tuning voltage. (c) Integrated intensity of QD1 and QD2 in the drop-port as a function of voltage. As the voltage is controlled, QD1 and QD2 follow the Lorentzian shape of the ring resonator transmission function.

Supplementary Notes

## Supplementary Note S.1

As discussed in the main manuscript, the waveguide dimensions (800 nm x 200 nm) and gap between the ring-waveguide and the bus-waveguide (180 nm) are all optimized to achieved critical coupling for TE resonant modes. The TM resonant modes are critically coupled at larger waveguide separations (~340 nm), while being over-coupled in the presented devices with 180 nm gaps ($Q_{coupling} < Q_{intrinsic}$). The through-port transmission of the TM resonant modes is shown in **Figure.S1**, the resonances show lower loaded quality factor than the TE modes due to the additional loss rate provided by the leakage to the bus-waveguides.

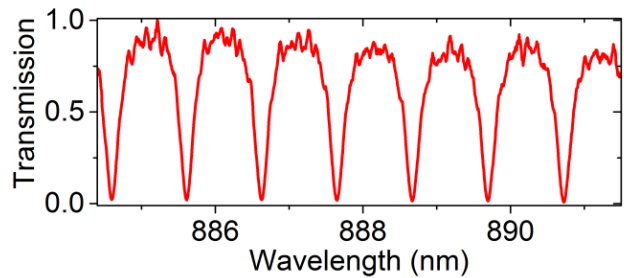

**Figure.S1** Through-port transmission of the TM mode in the ring resonator.

## Supplementary Note S.2

In order to accurately determine the QD temperature while tuning the ring resonator filter, we studied the QD emission wavelength shift as a function of the cryostat temperature. As shown in **Figure.S2**, the emission wavelength is a non-linear function of the temperature. By comparing the QD emission wavelength shift at a specific voltage with the wavelength shifts at a certain cryostat temperature, we can deduce the QD temperature while tuning the ring resonator filter. We estimated the QD temperature at 12.5 Volts of tuning voltage to be < 35 K.

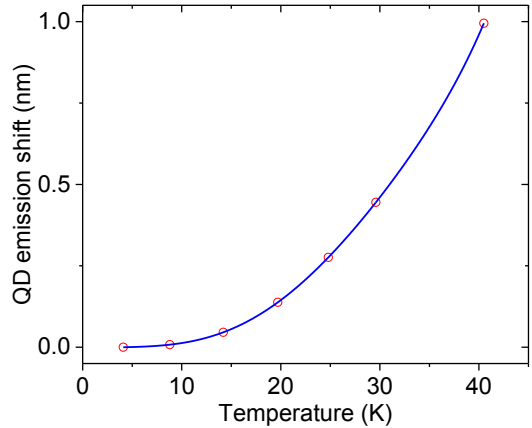

**Figure.S2** Emission wavelength shift of QD versus the cryostat temperature.

**Supplementary Note S.3**

The excellent on-chip filtering that is demonstrated can be attributed to a combination of attenuation and filtering of unwanted photons. Ultra-efficient pump-suppression in the drop port is a result of material absorption, severe under-coupling of the pump to the resonator, and the use of top excitation. The high-frequency chemical vapor deposition of SiN (1:1 $NH_3:SiH_4$) exhibits high optical absorption at the visible range of the spectrum[1]. Additionally, the pump photons have much lower coupling to the resonating modes of the cavity. The ring-waveguide gap of 180 nm was designed to achieve critical coupling of TE modes at a wavelength of ~880 nm. For the pump photons, the effective length of the two coupling gaps is increased by ~400 nm, resulting in close to zero transmission to the drop port. A similar argument is valid for the bulk InP wire emission at approximately 830 nm shown in Figure.S.3 with excitation power $P_{excitation} \gg P_{saturation}$. Therefore, the bulk InP emission couples less efficiently to the drop port.

Furthermore, since the waveguide dimensions impose a single mode cut-off for 532 nm photons, the pump is coupled to higher order modes which exhibit weak coupling to the resonant modes of the cavity due to the small overlap integral between the evanescent fields.

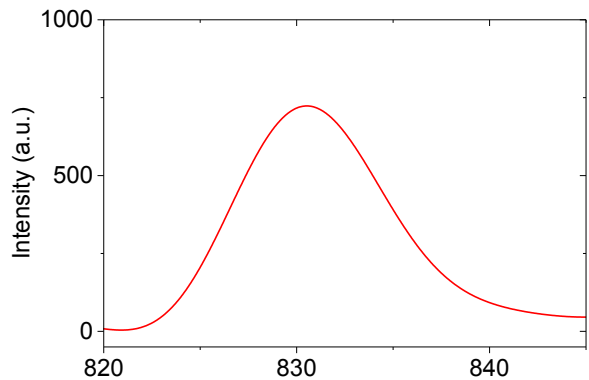

**Figure.S3** Nanowire bulk InP emission.

## Supplementary Note S.4

The slow detector response limited the resolution of the measured correlation functions. We independently measured the response time of the APD using a 3 picoseconds pulsed laser to be >350ps. By incorporating the response function of the detectors, we directly fit the measured the second-order correlation functions limited by the detector response (green line in **Figure.3d** and **Figure.3e** in the manuscript). The data fit yields $g^2(0) = 0.59 \pm 0.05$ for the ring resonator filtering and $g^2(0) = 0.39 \pm 0.04$ for the monochromator filtering.

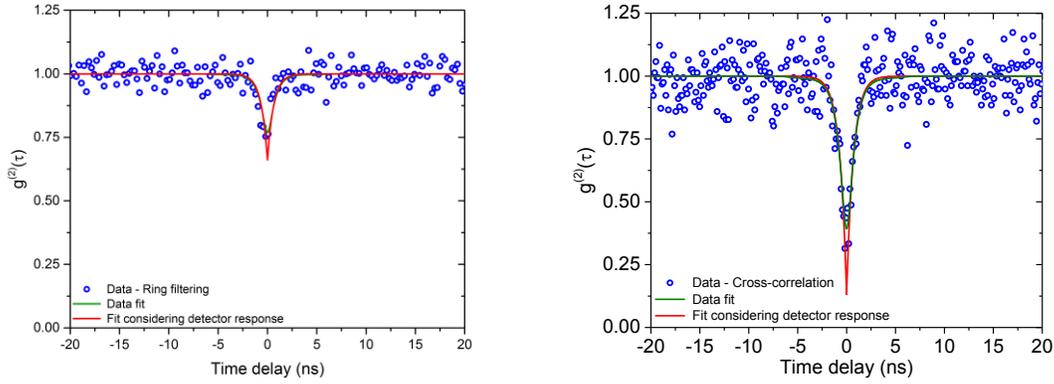

**Figure.S4** (a) Auto-correlation measurement with ring resonaor filtering considering both TE and TM modes at the same time, with QD emission aligned to TE resonance of the ring resonator. (b) cross-correlation measurement between the trion and exciton emission lines in **Figure.3b** in the manuscript.

**Figure.S4a** shows the second-order correlation function measured with only the ring resonator filtering, similar to **Figure.3d** in the main manuscript, but we consider both TE and TM modes at the same time. The multi-photon probability at zero time delay was measured to be $g^2(0) = 0.66 \pm 0.03$, considering the detector response. The value is slightly larger than the TE value presented in the main manuscript due to the finite contribution of the TM bulk InP nanowire emission through the over-coupled TM resonant modes. **Figure.S4b** shows cross-correlation measurement on the trion and exciton lines of the nanowire shown in **Figure.3b** in the main manuscript. The multi-photon probability at zero time delay was measured to be $g^2(0) = 0.13 \pm 0.06$ ($g^2(0) = 0.23 \pm 0.06$ without considering the detector response).

## Supplementary Note S.5

We performed finite difference time domain simulations to estimate the coupling efficiency between the butt-coupled nanowires and the photonic waveguide. The simulated structure and a horizontal cut depicting the electric field distribution of the TE mode are shown in **Figure.S5a** and **Figure.S5b**, respectively. The coupling efficiency to the waveguide was calculated to be 3% of the total forward emitted power in the nanowire. The resonators in the multiplexing circuits are designed to have smaller gaps to increase coupling to the drop port. Nanowire encapsulation and the use of 1D Bragg reflectors can be used to achieve a uni-directional emission with much higher efficiencies exceeding 90% [2].

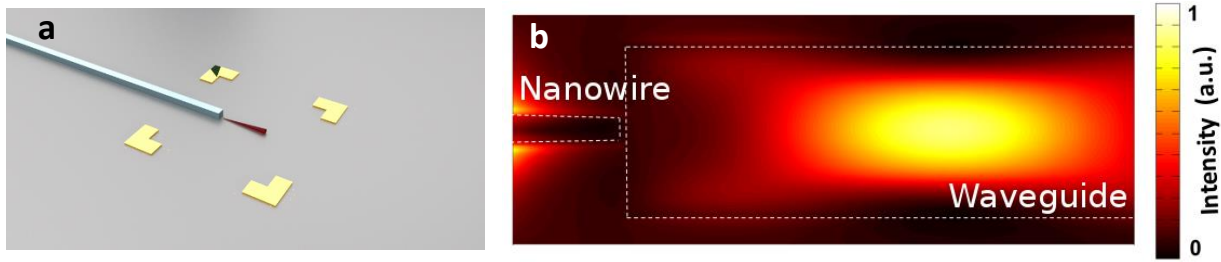

**Figure.S5** (a) Simulated device of a nanowire QD butt coupled to a photonic waveguide.(b) Electric field distribution of the TE mode emitted from nanowire QD to the photonic waveguide.

## Supplementary Note S.6

Here we present the results of another QWDM device. Two nanowires QDs are butt coupled to a SiN waveguide, their emission wavelengths are separated by ~6 nm. As shown in **Figure.S6,** we are able to wavelength-multiplex the emission of both QDs into one photonic channel, then demultiplexing the emission using a tunable filter. As we increase the tuning voltage, QD1 can be decoupled from the through-port of the ring resonator leaving only the signal from QD2. In this device, the voltages applied are higher than the QWDM device presented in the main manuscript, this is attributed to the fact that the we used an off-chip heater for tuning.

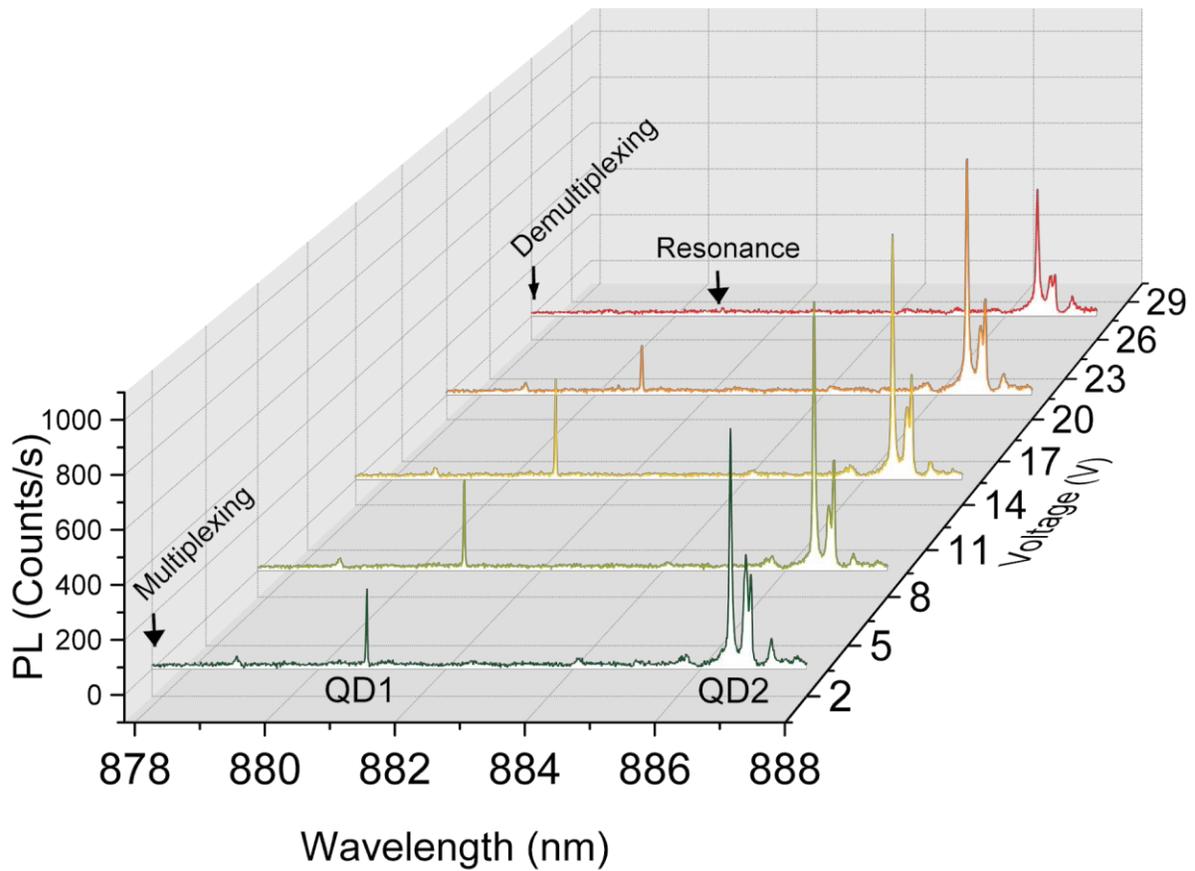

**Figure.S6** Through-port transmission of the photonic channel as a function of the heater voltage voltage. At low voltages, both QDs are wavelength multiplexed into a single photonic channel. As the voltage is increase, QD1 can be filtered-out from the through-port of the ring resonator leaving only the signal from QD2.

References :

[1] A. Gorin, A. Jaouad, E. Grondin, V. Aimez, P. Charette, Optics Express, 16 (2008) 13509-13516.
[2] I.E. Zadeh, A.W. Elshaari, K.D. Jöns, A. Fognini, D. Dalacu, P.J. Poole, M.E. Reimer, V. Zwiller, Nano Letters, 16 (2016) 2289-2294.